\documentclass[%
reprint,
superscriptaddress,
nofootinbib,
amsmath,amssymb,
aps
]{revtex4-2}

\usepackage{graphicx}
\usepackage{subcaption} 
\usepackage{dcolumn}
\usepackage{bm}

\usepackage{color}
\usepackage[table,xcdraw,dvipsnames]{xcolor}
\usepackage{orcidlink} 
\hypersetup{
    colorlinks,
    linkcolor={red!50!black},
    citecolor={blue!50!black},
    urlcolor={blue!80!black}
} 

\definecolor{blue-violet}{rgb}{0.7, 0.2, 0.8}

\renewcommand\footnotemark{}

\begin{document}

\preprint{APS/123-QED}

 \title{The second data release from the European Pulsar Timing Array\\VI. Challenging the ultralight dark matter paradigm
 }
\author{Clemente Smarra\orcidlink{0000-0002-0817-2830}}
 \email{csmarra@sissa.it}
\affiliation{SISSA — International School for Advanced Studies, Via Bonomea 265, 34136, Trieste, Italy and INFN, Sezione di Trieste}
\affiliation{IFPU — Institute for Fundamental Physics of the Universe, Via Beirut 2, 34014 Trieste, Italy}

\author{Boris Goncharov\orcidlink{0000-0003-3189-5807}}%
\affiliation{Gran Sasso Science Institute (GSSI), I-67100 L'Aquila, Italy}
\affiliation{INFN, Laboratori Nazionali del Gran Sasso, I-67100 Assergi, Italy}

\author{Enrico Barausse\orcidlink{0000-0001-6499-6263 }}
\affiliation{SISSA — International School for Advanced Studies, Via Bonomea 265, 34136, Trieste, Italy and INFN, Sezione di Trieste}
\affiliation{IFPU — Institute for Fundamental Physics of the Universe, Via Beirut 2, 34014 Trieste, Italy}

\author{J.~Antoniadis\orcidlink{0000-0003-4453-776}}
\affiliation{Institute of Astrophysics, FORTH, N. Plastira 100, 70013, Heraklion, Greece} 
\affiliation{Max-Planck-Institut f{\"u}r Radioastronomie, Auf dem H{\"u}gel 69, 53121 Bonn, Germany}

\author{S.~Babak\orcidlink{0000-0001-7469-4250}}
\affiliation{Universit{\'e} Paris Cit{\'e} CNRS, Astroparticule et Cosmologie, 75013 Paris, France}

\author{A.-S.~Bak~Nielsen\orcidlink{ 0000-0002-1298-9392}}
\affiliation{Max-Planck-Institut f{\"u}r Radioastronomie, Auf dem H{\"u}gel 69, 53121 Bonn, Germany}
\affiliation{Fakult{\"a}t f{\"u}r Physik, Universit{\"a}t Bielefeld, Postfach 100131, 33501 Bielefeld, Germany}

\author{C.~G.~Bassa\orcidlink{0000-0002-1429-9010}}
\affiliation{ASTRON, Netherlands Institute for Radio Astronomy, Oude Hoogeveensedijk 4, 7991 PD, Dwingeloo, The Netherlands}

\author{A.~Berthereau}
\affiliation{Laboratoire de Physique et Chimie de l'Environnement et de l'Espace, Universit\'e d'Orl\'eans / CNRS, 45071 Orl\'eans Cedex 02, France }
\affiliation{Observatoire Radioastronomique de Nan\c{c}ay, Observatoire de Paris, Universit\'e PSL, Université d'Orl\'eans, CNRS, 18330 Nan\c{c}ay, France}

\author{M.~Bonetti\orcidlink{0000-0001-7889-6810}}
\affiliation{Dipartimento di Fisica ``G. Occhialini", Universit{\'a} degli Studi di Milano-Bicocca, Piazza della Scienza 3, I-20126 Milano, Italy}
\affiliation{INFN, Sezione di Milano-Bicocca, Piazza della Scienza 3, I-20126 Milano, Italy}
\affiliation{INAF - Osservatorio Astronomico di Brera, via Brera 20, I-20121 Milano, Italy}

\author{E.~Bortolas}
\affiliation{Dipartimento di Fisica ``G. Occhialini", Universit{\'a} degli Studi di Milano-Bicocca, Piazza della Scienza 3, I-20126 Milano, Italy}
\affiliation{INFN, Sezione di Milano-Bicocca, Piazza della Scienza 3, I-20126 Milano, Italy}
\affiliation{INAF - Osservatorio Astronomico di Brera, via Brera 20, I-20121 Milano, Italy}

\author{P.~R.~Brook\orcidlink{0000-0003-3053-6538}}
\affiliation{Institute for Gravitational Wave Astronomy and School of Physics and Astronomy, University of Birmingham, Edgbaston, Birmingham B15 2TT, UK}

\author{M.~Burgay\orcidlink{0000-0002-8265-4344}}
\affiliation{INAF - Osservatorio Astronomico di Cagliari, via della Scienza 5, 09047 Selargius (CA), Italy}

\author{R.~N.~Caballero\orcidlink{0000-0001-9084-9427}}
\affiliation{Hellenic Open University, School of Science and Technology, 26335 Patras, Greece}

\author{A.~Chalumeau\orcidlink{0000-0003-2111-1001}}
\affiliation{Dipartimento di Fisica ``G. Occhialini", Universit{\'a} degli Studi di Milano-Bicocca, Piazza della Scienza 3, I-20126 Milano, Italy}

\author{D.~J.~Champion\orcidlink{0000-0003-1361-7723}}
\affiliation{Max-Planck-Institut f{\"u}r Radioastronomie, Auf dem H{\"u}gel 69, 53121 Bonn, Germany}

\author{S.~Chanlaridis\orcidlink{0000-0002-9323-9728}}
\affiliation{Institute of Astrophysics, FORTH, N. Plastira 100, 70013, Heraklion, Greece} 

\author{S.~Chen\orcidlink{0000-0002-3118-5963}}
\affiliation{Kavli Institute for Astronomy and Astrophysics, Peking University, Beijing 100871, P. R. China}

\author{I.~Cognard\orcidlink{0000-0002-1775-9692}}
\affiliation{Laboratoire de Physique et Chimie de l'Environnement et de l'Espace, Universit\'e d'Orl\'eans / CNRS, 45071 Orl\'eans Cedex 02, France }
\affiliation{Observatoire Radioastronomique de Nan\c{c}ay, Observatoire de Paris, Universit\'e PSL, Université d'Orl\'eans, CNRS, 18330 Nan\c{c}ay, France}

\author{G.~Desvignes\orcidlink{0000-0003-3922-4055}}
\affiliation{Max-Planck-Institut f{\"u}r Radioastronomie, Auf dem H{\"u}gel 69, 53121 Bonn, Germany}

\author{M.~Falxa}
\affiliation{Laboratoire de Physique et Chimie de l'Environnement et de l'Espace, Universit\'e d'Orl\'eans / CNRS, 45071 Orl\'eans Cedex 02, France }
\affiliation{Universit{\'e} Paris Cit{\'e} CNRS, Astroparticule et Cosmologie, 75013 Paris, France}

\author{R.~D.~Ferdman}
\affiliation{School of Physics, Faculty of Science, University of East Anglia, Norwich NR4 7TJ, UK}

\author{A.~Franchini\orcidlink{0000-0002-8400-0969}}
\affiliation{Dipartimento di Fisica ``G. Occhialini", Universit{\'a} degli Studi di Milano-Bicocca, Piazza della Scienza 3, I-20126 Milano, Italy}
\affiliation{INFN, Sezione di Milano-Bicocca, Piazza della Scienza 3, I-20126 Milano, Italy}

\author{J.~R.~Gair\orcidlink{0000-0002-1671-3668}}
\affiliation{Max Planck Institute for Gravitational Physics (Albert Einstein Institute), Am Mu{\"u}hlenberg 1, 14476 Potsdam, Germany}

\author{E.~Graikou}
\affiliation{Max-Planck-Institut f{\"u}r Radioastronomie, Auf dem H{\"u}gel 69, 53121 Bonn, Germany}

\author{J.-M.~Grie{\ss}meier\orcidlink{0000-0003-3362-7996}}
\affiliation{Laboratoire de Physique et Chimie de l'Environnement et de l'Espace, Universit\'e d'Orl\'eans / CNRS, 45071 Orl\'eans Cedex 02, France }
\affiliation{Observatoire Radioastronomique de Nan\c{c}ay, Observatoire de Paris, Universit\'e PSL, Université d'Orl\'eans, CNRS, 18330 Nan\c{c}ay, France}

\author{L.~Guillemot\orcidlink{0000-0002-9049-8716}}
\affiliation{Laboratoire de Physique et Chimie de l'Environnement et de l'Espace, Universit\'e d'Orl\'eans / CNRS, 45071 Orl\'eans Cedex 02, France }
\affiliation{Observatoire Radioastronomique de Nan\c{c}ay, Observatoire de Paris, Universit\'e PSL, Université d'Orl\'eans, CNRS, 18330 Nan\c{c}ay, France}

\author{Y.~J.~Guo}
\affiliation{Max-Planck-Institut f{\"u}r Radioastronomie, Auf dem H{\"u}gel 69, 53121 Bonn, Germany}

\author{H.~Hu\orcidlink{0000-0002-3407-8071}}
\affiliation{Max-Planck-Institut f{\"u}r Radioastronomie, Auf dem H{\"u}gel 69, 53121 Bonn, Germany}

\author{F.~Iraci}
\affiliation{INAF - Osservatorio Astronomico di Cagliari, via della Scienza 5, 09047 Selargius (CA), Italy}
\affiliation{Universit{\'a} di Cagliari, Dipartimento di Fisica, S.P. Monserrato-Sestu Km 0,700 - 09042 Monserrato (CA), Italy}

\author{D.~Izquierdo-Villalba\orcidlink{0000-0002-6143-1491}}
\affiliation{Dipartimento di Fisica ``G. Occhialini", Universit{\'a} degli Studi di Milano-Bicocca, Piazza della Scienza 3, I-20126 Milano, Italy}
\affiliation{INFN, Sezione di Milano-Bicocca, Piazza della Scienza 3, I-20126 Milano, Italy}

\author{J.~Jang\orcidlink{0000-0003-4454-0204}}
\affiliation{Max-Planck-Institut f{\"u}r Radioastronomie, Auf dem H{\"u}gel 69, 53121 Bonn, Germany}

\author{J.~Jawor\orcidlink{0000-0003-3391-0011}}
\affiliation{Max-Planck-Institut f{\"u}r Radioastronomie, Auf dem H{\"u}gel 69, 53121 Bonn, Germany}

\author{G.~H.~Janssen\orcidlink{0000-0003-3068-3677}}
\affiliation{ASTRON, Netherlands Institute for Radio Astronomy, Oude Hoogeveensedijk 4, 7991 PD, Dwingeloo, The Netherlands}
\affiliation{Department of Astrophysics/IMAPP, Radboud University Nijmegen, P.O. Box 9010, 6500 GL Nijmegen, The Netherlands}

\author{A.~Jessner}
\affiliation{Max-Planck-Institut f{\"u}r Radioastronomie, Auf dem H{\"u}gel 69, 53121 Bonn, Germany}

\author{R.~Karuppusamy\orcidlink{0000-0002-5307-2919}}
\affiliation{Max-Planck-Institut f{\"u}r Radioastronomie, Auf dem H{\"u}gel 69, 53121 Bonn, Germany}

\author{E.~F.~Keane\orcidlink{0000-0002-4553-655X}}
\affiliation{School of Physics, Trinity College Dublin, College Green, Dublin 2, D02 PN40, Ireland}

\author{M.~J.~Keith\orcidlink{0000-0001-5567-5492}}
\affiliation{Jodrell Bank Centre for Astrophysics, Department of Physics and Astronomy, University of Manchester, Manchester M13 9PL, UK}

\author{M.~Kramer}
\affiliation{Max-Planck-Institut f{\"u}r Radioastronomie, Auf dem H{\"u}gel 69, 53121 Bonn, Germany}
\affiliation{Jodrell Bank Centre for Astrophysics, Department of Physics and Astronomy, University of Manchester, Manchester M13 9PL, UK}

\author{M.~A.~Krishnakumar\orcidlink{0000-0003-4528-2745}}
\affiliation{Max-Planck-Institut f{\"u}r Radioastronomie, Auf dem H{\"u}gel 69, 53121 Bonn, Germany}
\affiliation{Fakult{\"a}t f{\"u}r Physik, Universit{\"a}t Bielefeld, Postfach 100131, 33501 Bielefeld, Germany}

\author{K.~Lackeos\orcidlink{0000-0002-6554-3722}}
\affiliation{Max-Planck-Institut f{\"u}r Radioastronomie, Auf dem H{\"u}gel 69, 53121 Bonn, Germany}

\author{K.~J.~Lee}
\affiliation{Institute of Astrophysics, FORTH, N. Plastira 100, 70013, Heraklion, Greece} 
\affiliation{Max-Planck-Institut f{\"u}r Radioastronomie, Auf dem H{\"u}gel 69, 53121 Bonn, Germany}
\affiliation{Observatoire Radioastronomique de Nan\c{c}ay, Observatoire de Paris, Universit\'e PSL, Université d'Orl\'eans, CNRS, 18330 Nan\c{c}ay, France}

\author{K.~Liu}
\affiliation{Max-Planck-Institut f{\"u}r Radioastronomie, Auf dem H{\"u}gel 69, 53121 Bonn, Germany}

\author{Y.~Liu\orcidlink{0000-0001-9986-9360}}
\affiliation{Fakult{\"a}t f{\"u}r Physik, Universit{\"a}t Bielefeld, Postfach 100131, 33501 Bielefeld, Germany}
\affiliation{National Astronomical Observatories, Chinese Academy of Sciences, Beijing 100101, P. R. China}

\author{A.~G.~Lyne}
\affiliation{Jodrell Bank Centre for Astrophysics, Department of Physics and Astronomy, University of Manchester, Manchester M13 9PL, UK}

\author{J.~W.~McKee\orcidlink{0000-0002-2885-8485}}
\affiliation{E.A. Milne Centre for Astrophysics, University of Hull, Cottingham Road, Kingston-upon-Hull, HU6 7RX, UK}
\affiliation{Centre of Excellence for Data Science, Artificial Intelligence and Modelling (DAIM), University of Hull, Cottingham Road, Kingston-upon-Hull, HU6 7RX, UK}

\author{R.~A.~Main}
\affiliation{Max-Planck-Institut f{\"u}r Radioastronomie, Auf dem H{\"u}gel 69, 53121 Bonn, Germany}

\author{M.~B.~Mickaliger\orcidlink{0000-0001-6798-5682}}
\affiliation{Jodrell Bank Centre for Astrophysics, Department of Physics and Astronomy, University of Manchester, Manchester M13 9PL, UK}

\author{I.~C.~Ni\c{t}u\orcidlink{0000-0003-3611-3464}}
\affiliation{Jodrell Bank Centre for Astrophysics, Department of Physics and Astronomy, University of Manchester, Manchester M13 9PL, UK}

\author{A.~Parthasarathy\orcidlink{0000-0002-4140-5616}}
\affiliation{Max-Planck-Institut f{\"u}r Radioastronomie, Auf dem H{\"u}gel 69, 53121 Bonn, Germany}

\author{B.~B.~P.~Perera\orcidlink{0000-0002-8509-5947}}
\affiliation{Arecibo Observatory, HC3 Box 53995, Arecibo, PR 00612, USA}

\author{D.~Perrodin\orcidlink{0000-0002-1806-2483}}
\affiliation{INAF - Osservatorio Astronomico di Cagliari, via della Scienza 5, 09047 Selargius (CA), Italy}

\author{A.~Petiteau\orcidlink{0000-0002-7371-9695}}
\affiliation{IRFU, CEA, Université Paris-Saclay, F-91191 Gif-sur-Yvette, France}
\affiliation{Universit{\'e} Paris Cit{\'e} CNRS, Astroparticule et Cosmologie, 75013 Paris, France}

\author{N.~K.~Porayko}
\affiliation{Max-Planck-Institut f{\"u}r Radioastronomie, Auf dem H{\"u}gel 69, 53121 Bonn, Germany}
\affiliation{Dipartimento di Fisica ``G. Occhialini", Universit{\'a} degli Studi di Milano-Bicocca, Piazza della Scienza 3, I-20126 Milano, Italy}

\author{A.~Possenti}
\affiliation{INAF - Osservatorio Astronomico di Cagliari, via della Scienza 5, 09047 Selargius (CA), Italy}

\author{H.~Quelquejay~Leclere\orcidlink{0000-0002-6766-2004}}
\affiliation{Universit{\'e} Paris Cit{\'e} CNRS, Astroparticule et Cosmologie, 75013 Paris, France}

\author{A.~Samajdar\orcidlink{0000-0002-0857-6018}}
\affiliation{Institut f\"{u}r Physik und Astronomie, Universit\"{a}t Potsdam, Haus 28, Karl-Liebknecht-Str. 24/25, 14476, Potsdam, Germany}

\author{S.~A.~Sanidas}
\affiliation{Jodrell Bank Centre for Astrophysics, Department of Physics and Astronomy, University of Manchester, Manchester M13 9PL, UK}

\author{A.~Sesana}
\affiliation{Dipartimento di Fisica ``G. Occhialini", Universit{\'a} degli Studi di Milano-Bicocca, Piazza della Scienza 3, I-20126 Milano, Italy}
\affiliation{INFN, Sezione di Milano-Bicocca, Piazza della Scienza 3, I-20126 Milano, Italy}
\affiliation{INAF - Osservatorio Astronomico di Brera, via Brera 20, I-20121 Milano, Italy}

\author{G.~Shaifullah\orcidlink{0000-0002-8452-4834}}
\affiliation{Dipartimento di Fisica ``G. Occhialini", Universit{\'a} degli Studi di Milano-Bicocca, Piazza della Scienza 3, I-20126 Milano, Italy}
\affiliation{INFN, Sezione di Milano-Bicocca, Piazza della Scienza 3, I-20126 Milano, Italy}
\affiliation{INAF - Osservatorio Astronomico di Cagliari, via della Scienza 5, 09047 Selargius (CA), Italy}

\author{L.~Speri\orcidlink{0000-0002-5442-7267}}
\affiliation{Max Planck Institute for Gravitational Physics (Albert Einstein Institute), Am Mu{\"u}hlenberg 1, 14476 Potsdam, Germany}

\author{R.~Spiewak}
\affiliation{Jodrell Bank Centre for Astrophysics, Department of Physics and Astronomy, University of Manchester, Manchester M13 9PL, UK}

\author{B.~W.~Stappers}
\affiliation{Jodrell Bank Centre for Astrophysics, Department of Physics and Astronomy, University of Manchester, Manchester M13 9PL, UK}

\author{S.~C.~Susarla\orcidlink{0000-0003-4332-8201}}
\affiliation{Ollscoil na Gaillimhe --- University of Galway, University Road, Galway, H91 TK33, Ireland}

\author{G.~Theureau\orcidlink{0000-0002-3649-276X}}
\affiliation{Laboratoire de Physique et Chimie de l'Environnement et de l'Espace, Universit\'e d'Orl\'eans / CNRS, 45071 Orl\'eans Cedex 02, France }
\affiliation{Observatoire Radioastronomique de Nan\c{c}ay, Observatoire de Paris, Universit\'e PSL, Université d'Orl\'eans, CNRS, 18330 Nan\c{c}ay, France}
\affiliation{Laboratoire Univers et Th{\'e}ories LUTh, Observatoire de Paris, Universit{\'e} PSL, CNRS, Universit{\'e} de Paris, 92190 Meudon, France}

\author{C.~Tiburzi}
\affiliation{INAF - Osservatorio Astronomico di Cagliari, via della Scienza 5, 09047 Selargius (CA), Italy}

\author{E.~van~der~Wateren\orcidlink{0000-0003-0382-8463}}
\affiliation{Department of Astrophysics/IMAPP, Radboud University Nijmegen, P.O. Box 9010, 6500 GL Nijmegen, The Netherlands}
\affiliation{ASTRON, Netherlands Institute for Radio Astronomy, Oude Hoogeveensedijk 4, 7991 PD, Dwingeloo, The Netherlands}

\author{A.~Vecchio\orcidlink{0000-0002-6254-1617}}
\affiliation{Institute for Gravitational Wave Astronomy and School of Physics and Astronomy, University of Birmingham, Edgbaston, Birmingham B15 2TT, UK}

\author{V.~Venkatraman~Krishnan\orcidlink{0000-0001-9518-9819}}
\affiliation{Max-Planck-Institut f{\"u}r Radioastronomie, Auf dem H{\"u}gel 69, 53121 Bonn, Germany}

\author{J.~Wang\orcidlink{0000-0003-1933-6498}}
\affiliation{Fakult{\"a}t f{\"u}r Physik, Universit{\"a}t Bielefeld, Postfach 100131, 33501 Bielefeld, Germany}
\affiliation{Ruhr University Bochum, Faculty of Physics and Astronomy, Astronomical Institute (AIRUB), 44780 Bochum, Germany}
\affiliation{Advanced Institute of Natural Sciences, Beijing Normal University, Zhuhai 519087, China }

\author{L.~Wang}
\affiliation{Jodrell Bank Centre for Astrophysics, Department of Physics and Astronomy, University of Manchester, Manchester M13 9PL, UK}

\author{Z.~Wu\orcidlink{0000-0002-1381-7859}}
\affiliation{National Astronomical Observatories, Chinese Academy of Sciences, Beijing 100101, P. R. China}

\collaboration{The European Pulsar Timing Array}

\date{\today}

\begin{abstract}

Pulsar Timing Array experiments probe the presence of possible scalar or pseudoscalar ultralight dark matter particles through decade-long timing of an ensemble of galactic millisecond radio pulsars. With the second data release of the European Pulsar Timing Array, we focus on the most robust scenario, in which dark matter interacts only gravitationally with ordinary baryonic matter. 
Our results show that ultralight particles with masses $10^{-24.0}~\text{eV} \lesssim m \lesssim 10^{-23.3}~\text{eV}$ cannot constitute $100\%$ of the measured local dark matter density, but can have at most local density $\rho\lesssim 0.3$ GeV/cm$^3$. 

\end{abstract}



\maketitle


\textit{Introduction.}---The nature of Dark Matter (DM) is arguably one of the most fascinating and mysterious questions that we are struggling to answer. Galaxy rotation curves~\cite{Rubin_1970, Rubin_1980}, the peculiar motion of clusters~\cite{Zwicky_1933, Zwicky_1937}, the Bullet Cluster system~\cite{Clowe_2006}, measurements of cosmological abundances from Cosmic Microwave Background (CMB) and Baryonic Acoustic Oscillation (BAO) observations~\cite{Planck_2020, Bennett_2013}  all point to the existence of a hitherto-unseen type of matter, constituting roughly 26\% of the current energy density of the Universe and interacting mostly gravitationally with the Standard Model of particle physics. 
The standard Cold Dark Matter (CDM) paradigm describes successfully many aspects of the large-scale structure of the Universe, but struggles to predict what we observe at scales smaller than the $\sim$kpc. For instance, observations favor a constant density profile in the inner part of  galaxies, while CDM predicts a steep power-law-like behavior (\textit{cusp-core problem}) \cite{Flores_1994, Moore_1994, Karukes_2015}. Furthermore, well-known issues are associated with the discrepancy between the observed and expected number of Milky Way (MW) satellites (\textit{missing satellite problem})~\cite{Klypin_1999, Moore_1999} and with $\Lambda\text{CDM}$ simulations showing that the most massive subhaloes of the MW would be too dense to host any of its bright satellites (\textit{too-big-to-fail problem})~\cite{Boylan-Kolchin_2011}. Moreover, recent anomalies in gravitationally lensed images~\cite{Amruth_2023} seem to disfavor the long-standing Weakly-Interacting-Massive-Particles (WIMPs) hypothesis for CDM. While some of these issues might be alleviated by invoking baryonic feedback mechanisms \cite{DeLaurentis_2022}, e.g. Active Galactic Nuclei (AGN)~\cite{Morganti_2017} and/or supernova feedback~\cite{Navarro_1996, Governato_2012, Brooks_2013, Chan_2015, Onorbe_2015, Read_2016},  it is still unclear how the flat density profile of dwarf galaxies (e.g. Fornax~\cite{Jardel_2012}), with almost no baryonic activity in the center, can be explained without invoking a novel mechanism. An intriguing alternative is to consider the possibility that DM is fuzzy, \textit{i.e.} an ultra-light scalar field ($m_\phi \sim 10^{-22}~\text{eV}$) or axion-like particle, whose wavelike nature suppresses structure formation on scales smaller than the de Broglie wavelength, while maintaining all the achievements of the CDM paradigm on large scales. 
Moreover, the existence of ultralight scalars can also be motivated on a more theoretical ground: in particular, axion-like particles generically arise in string theory compactifications as Kaluza–Klein zero modes of antisymmetric tensor fields~\cite{Green_1987, Svrcek_2006, Arvanitaki_2010}.

A wealth of studies have been carried out to probe the existence of ultra-light dark matter (ULDM), ranging from CMB observables to Lyman-$\alpha$ and stellar kinematics. Specifically, the integrated Sachs-Wolfe effect on CMB anisotropies rules out masses $m_\phi \lesssim 10^{-24}~\text{eV}$ \cite{Hlozek_2015}, while Lyman-$\alpha$ gives a lower bound $m_\phi \gtrsim 10^{-21}~\text{eV}$ for ultra-light candidates constituting more than $\sim 30\%$ of DM\footnote{Notice, however, that the value of the fraction of ULDM compatible with Lyman-$\alpha$ is extrapolated for masses $m_\phi\lesssim 10^{-22}~\text{eV}$.}~\cite{Irsic_2017,Armengaud_2017,Kobayashi_2017,Nori_2018, Rogers_2021}. Stellar orbit kinematics in ultra-faint dwarf (UFD) galaxies might even be able to bound the scalar field mass to be $m_\phi \gtrsim 10^{-19}~\text{eV}$, although this  is still under debate~\cite{Hayashi_2021, Dalal_2022}. However, the  sensitivity of non-CMB constraints to uncertainties in the modeling of small scale structure properties~\cite{Schive_2014, Zhang_2019} makes it compelling to rely on complementary and independent probes.
It was shown by Khmelnitsky and Rubakov~\cite{Khmelnitsky_2014} that the presence of ULDM induces an oscillating gravitational potential that affects the light travel time of radio pulses emitted by pulsars. Therefore, Pulsar Timing Arrays (PTAs) stand out as promising experiments to test the presence of ULDM particles in the MW.
Previous PTA searches placed 95\% upper limits on the local energy density of ULDM at $3 \times 10^{-24}$~eV to $\lesssim 1~\text{GeV}/\text{cm}^{3}$~\cite{Porayko_2014, Porayko_2018, Xia_2023}.

In this work, which is complementary to the European Pulsar Timing Array (EPTA) interpretation effort~\cite{epta_interpret}, we 
focus on a specific range of ULDM masses
and constrain the local ULDM density to values
\textit{below} the observed local DM density. We do so by analyzing the effect of ULDM on the times of arrival (TOAs) of pulsar radio beams. Therefore, if ULDM particles exist in the mass range that we consider, they cannot constitute all of the observed DM.

\textit{Models.}---As we only have gravitational evidence of DM, we focus on an ultra-light scalar field with negligible self-interactions and no couplings with the Standard Model. The action for this field can be written as 
\begin{equation}
    S=\int d^4x\sqrt{-g}\left[\frac{1}{2}g^{\mu\nu}\partial_\mu\phi\partial_\nu\phi-\frac{1}{2}m_\phi^2\phi^2\right]\,.
    \label{eq:action}
\end{equation}
Due to its high occupation number and  non-relativistic nature, the ULDM scalar field can be thought as a classical wave~\cite{Khmelnitsky_2014}:
\begin{equation}
\phi(\vec{x}, t)=\frac{\sqrt{2 \rho_\phi}}{m_\phi} \hat{\phi}(\vec{x}) \cos \left(m_\phi t+\gamma(\vec{x})\right),
\label{eq:phi}
\end{equation}
where  
$m_\phi$ is the mass of the scalar field, $\gamma(\vec{x})$ is a space-dependent phase and $\hat{\phi}(\vec{x})$ accounts for the pattern of interference in the proximity of $\vec{x}$ caused by the wave-like nature of ULDM. 
The scalar field density $\rho_\phi$ is conveniently normalized to the local DM density $\rho_{\text{DM}}$, which can be determined e.g. by fitting the MW rotation curve or, in a more refined way, by studying the vertical oscillations of disc stars~\cite{Bovy_2012,Read_2014,Sivertsson_2018,de_Salas_2020}.
In the following, we assume a fiducial value $\rho_{\text{DM}} \approx 0.4~\text{GeV}/\text{cm}^3$.
The oscillating nature of ULDM induces an oscillating gravitational potential leading to a periodic displacement $\delta t_\text{DM}$ in the TOAs of radio pulses emitted by pulsars, which can be written as~\cite{Khmelnitsky_2014, Porayko_2018}:
\begin{equation}
    \delta t_\text{DM} = \frac{\Psi_\text{c}(\vec{x})}{2m_\phi} [\hat{\phi}^2_\text{E}\sin{(2m_\phi + \gamma_\text{E} )} - \hat{\phi}^2_\text{P}\sin{(2m_\phi + \gamma_\text{P} )} ], 
    \label{eq:st}
\end{equation} 
where
\begin{equation}
    \frac{\Psi_\text{c}(\vec{x})}{10^{-18}} \approx 6.52 \left(\frac{10^{-22~}\text{eV}}{m_\phi}\right)^2 \left(\frac{\rho_\phi}{0.4~\text{GeV}/\text{cm}^3}\right),
\label{eq:psi_c}
\end{equation}
and $\gamma_\text{P} \equiv 2\gamma(\vec{x_\text{p}}) - 2 m_\phi d_p/c$ ($\gamma_\text{E} \equiv 2\gamma(\vec{x_\text{e}})$) are related to the phases of Eq.~\eqref{eq:phi} evaluated at the pulsar (Earth) location, with $d_p$ standing for the pulsar-Earth distance. 
The amplitude in Eq.~\eqref{eq:psi_c} is computed assuming a constant DM density background and possible deviations caused by the wave-like nature of the ultralight scalar field are parametrized in terms of the pulsar (Earth) dependent phase factors $\hat{\phi}^2(\vec{x_\text{p}})\equiv \hat{\phi}^2_\text{P} $ ($\hat{\phi}^2(\vec{x_\text{e}}) \equiv \hat{\phi}^2_\text{E}$). The approximation of constant DM density across pulsars is sufficient, as their distances from Earth are all $\sim$kpc and subject to large measurement uncertainties~\cite{Porayko_2018}. 
Notice that accurate measurements of pulsar-Earth distances might help to reduce the number of free parameters in the limit in which $\gamma(\vec{x_\text{p}}) = \gamma(\vec{x_\text{e}})$.
Moreover, precise determination of pulsar positions could provide us with more information about the dark matter density in its surroundings. 
On scales smaller than the de Broglie wavelength, the ULDM scalar field oscillates coherently, with the same amplitude $\hat\phi$ (see Eq.~\eqref{eq:phi}).
Since the typical ULDM velocity is expected to be $v_\phi \sim 10^{-3}$, the coherence length is approximately
\begin{equation}
    l_c \approx \frac{2\pi}{m_\phi v_\phi} \approx 0.4~\text{kpc} \left( \frac{10^{-22}~\text{eV}}{m_\phi} \right) \,.
    \label{eq:co_length}
\end{equation}
Therefore,  $\hat{\phi}^2_\text{E}$ and $\hat{\phi}^2_\text{P}$ are: 
\begin{itemize}
    \item \textit{uncorrelated} if the coherence length of ULDM is less than the average inter-pulsar and pulsar-Earth separation. In this case, $\hat{\phi}^2_\text{E}$ and $\hat{\phi}^2_\text{P}$ will thus be separate parameters;
    \item \textit{correlated} if the coherence length of ULDM is larger than the inter-pulsar and pulsar-Earth separations and encloses the typical Galacto-centric region tested by the most precise MW rotation curves measurements (roughly the inner $\sim 20$ kpc \cite{Nesti_2013}). In this case, $\hat{\phi}^2_\text{E} = \hat{\phi}^2_\text{P}$ for all the pulsars. Moreover, rotation curves 
    also sample from the same coherence patch,   and thus  
    measure the local abundance $\rho_{\text{DM}}$ of DM. Therefore, the stochastic parameter $\hat{\phi}^2$ can be safely absorbed in a redefinition of $\Psi_\text{c}$.
    \item \textit{pulsar-correlated} if the coherence length of ULDM is larger than the inter-pulsar and pulsar-Earth separations, but smaller than the typical Galacto-centric radius sampled by rotation curves. In this case, $\hat{\phi}^2_\text{E} = \hat{\phi}^2_\text{P}$ for all the pulsars. However, DM density estimates from rotation curves average over different patches. We therefore  keep $\hat{\phi}^2$ as a free parameter and consistently marginalize over it. In this way, the limits on $\rho_{\text{DM}}$ obtained from pulsars will constrain the same quantity measured by rotation curves.
\end{itemize}
We perform the analysis in the three limits above, noting that the fully correlated limit has not been considered in the previous studies~\cite{Kaplan_2022,Xue_2022}.
From Eq.~\eqref{eq:co_length}, recalling that the pulsar-Earth distance is $\mathcal{O}$(kpc) for the observed systems, it follows that the \textit{correlated} regime is an excellent approximation for masses lower than $m_\phi \sim 2 \times 10^{-24}$ eV; the \textit{pulsar-correlated} regime holds for $ 2 \times 10^{-24}\, \text{eV} \lesssim m_\phi \lesssim 5 \times 10^{-23}\, \text{eV} $ and the \textit{uncorrelated} regime is valid for $m_\phi \gtrsim 5\times 10^{-23}~\text{eV}$.
\\
\textit{Dataset and methodology.}---The EPTA monitors 42 millisecond radio pulsars with five telescopes located in France, Germany, Italy, the Netherlands, and the United Kingdom.
The second data release (DR2) of the EPTA contains 24.7 years of observations of pulse arrival times of 25 pulsars, surveyed with an approximate cadence of once every 3 weeks~\cite{epta_wm1}, which translates into a Nyquist frequency of approximately $3 \times 10^{-7}~\text{Hz}$.
TOAs are measured at the position of the Solar System Barycenter, and are primarily described by pulsar-specific deterministic timing models accounting for the position of each pulsar in the sky, spin down, proper motion, the presence of a binary companion, etc.
The timing models are provided with the data set.
Deviations between the TOAs predicted by these models and the measured TOAs are referred to as the timing residuals, $\delta t$.
The residuals contain contributions from various sources of noise, from variations in the dispersion measure to irregularities in pulsar rotation, but they may also contain  signals of astrophysical interest.
The sources of noise are identified as part of the noise analysis of the EPTA DR2~\cite{epta_wm2}.
The DR2 data set also hints at growing evidence for the stochastic gravitational-wave background, which manifests itself as a temporally-correlated stochastic process with a hallmark inter-pulsar correlation signature of General Relativity~\cite{epta_wm3}.
Our work is complementary to the EPTA-wide interpretation effort of the spatial and temporal correlations in DR2~\cite{epta_interpret}.

\begin{table*}[ht]
\renewcommand{\arraystretch}{1.2}
\centering
\caption{Parameters employed for the search along with their respective priors. In the correlated limit, the parameters $\hat\phi^2_\text{E}, \hat\phi^2_\text{P}$ are accounted for by a redefinition of $\Psi_c$, while in the pulsar-correlated regime $\hat\phi^2_\text{E} = \hat\phi^2_\text{P} = \hat\phi^2$ is a free parameter.   }
\begin{tabular}{|c|c|c|c|}
\hline  \textbf{Parameter} & \textbf{Description} & \textbf{Prior} & \textbf{Occurrence} \\ \hline
\hline \multicolumn{4}{|c|}{ White Noise $\left(\sigma = E_\text{f}^2 \sigma^2_\text{TOA} + E_\text{q}^2\right)$} \\ \hline
\hline$E_\text{f}$ & EFAC per receiver-backend system & Uniform $[0,10]$ & 1 per pulsar \\
\hline$E_\text{q}$ & EQUAD per receiver-backend system & Log$_{10}$-Uniform $[-10,-5]$ & 1 per pulsar \\  \hline
\hline \multicolumn{4}{|c|}{ Red Noise } \\  \hline
\hline$A_{\text {red }}$ & Red noise power-law amplitude & Log$_{10}$-Uniform $[-20,-6]$ & 1 per pulsar \\
\hline$\gamma_{\text {red }}$ & Red noise power-law spectral index & Uniform $[0,10]$ & 1 per pulsar \\  \hline
\hline \multicolumn{4}{|c|}{ ULDM } \\  \hline
\hline$\Psi_\text{c}$ & ULDM signal amplitude & $\text{Log}_{10}$-Uniform $[-20,-12]$ & 1 per PTA \\
\hline$m_\phi~[\mathrm{eV}]$ & ULDM mass & Log$_{10}$-Uniform $[-24,-22]$ & 1 per PTA \\
\hline$\hat{\phi}_\text{E}^2$ & Earth factor & $e^{-x}$ & 1 per PTA \\
\hline$\hat{\phi}_\text{P}^2$ & Pulsar factor & $e^{-x}$ & 1 per pulsar \\
\hline$\gamma_\text{E}$ & Earth signal phase & Uniform $[0,2 \pi]$ & 1 per PTA \\
\hline$\gamma_\text{P}$ & Pulsar signal phase & Uniform $[0,2 \pi]$ & 1 per pulsar \\  \hline
\hline \multicolumn{4}{|c|}{ Common spatially Uncorrelated Red Noise (CURN) } \\  \hline
\hline$A_{\mathrm{GWB}}$ & Common process strain amplitude & Log$_{10}$-Uniform $[-20,-6]$ & 1 per PTA \\
\hline
\end{tabular}
\label{tab:priors}
\end{table*}

We use 
Bayesian inference techniques 
to search for the ULDM signal while simultaneously fitting timing model parameters and all known sources of noise to the data, to correctly marginalize over the associated uncertainties. 
The likelihood of the timing residuals, $\mathcal{L}(\delta t|\theta)$, given the parameters of the models, $\theta$, is~\cite{vanHaasterenLevin2009,Lentati_2014, Taylor_2017,enterprise,enterprise_ext}
\begin{equation}
    \ln \mathcal{L}(\delta t | \theta) \propto  -\frac{1}{2} (\delta t - \mu)^\text{T} C^{-1}(\delta t-\mu).
\end{equation}
This is a time-domain Gaussian likelihood, multivariate with respect to
a number of observations, \textit{i.e.}, $\delta t$ has dimension equal to the number of
observations.
The contribution of ULDM from Eq.~\eqref{eq:st} 
is added to $\mu$, which also contains contributions from the timing model~\cite{epta_wm1} and noise processes, according to the noise analysis of Ref.~\cite{epta_wm2}.
The diagonal part of the covariance matrix $C$ contains TOA measurement uncertainties that include temporally-uncorrelated ``white'' noise.
Contributions from temporally correlated ``red'' noise may be added as off-diagonal elements in $C$.
However, for computational efficiency, red noise contributions are modeled in $\mu$~\cite{Lentati_2014,Taylor_2017}.
The priors $\pi(\theta)$ are set based on Table~\ref{tab:priors}, see Ref.~\cite{Kaplan_2022} for further details. 
To obtain a sufficient amount of posterior samples across the mass-frequency parameter space, the search is effectively performed across equally-spaced segments of $\pi(m_\phi)$, which we refer to as bins. 
The measurements of parameters are obtained as posterior distributions, $\mathcal{P}(\theta|d) \propto \mathcal{L}(\delta t|\theta) \pi(\theta)$.
The posteriors are evaluated using the Parallel-Tempering-Markov-Chain Monte-Carlo sampler~\cite{justin_ellis_2017_1037579} implemented in \textsc{enterprise}~\cite{enterprise} and \textsc{enterprise\_extensions}~\cite{enterprise_ext} .
The conclusion about the presence or absence of the ULDM signal in the data is based on the Bayesian odds ratio.
In our case, that is equal to the Bayes factor, because we assume the prior odds of both scenarios  to be equal.
We evaluate Bayes factors, $\mathcal{B}$, using the Savage-Dickey density ratio~\cite{dickey1971}. In particular, finding $\ln\mathcal{B} \gtrsim 5$ would indicate robust evidence for the ULDM signal.

There is strong evidence for a temporally-correlated red signal in EPTA DR2, characterized by the same Fourier spectrum of $\delta t$ in all pulsars~\cite{epta_wm3}.
This signal may contain contributions from the stochastic gravitational-wave background~\cite{GoncharovThrane2022}.
Because in the 24.7-yr data set
this signal does not show significant evidence for
inter-pulsar correlations (unlike in the 10.3-yr data set~\cite{epta_wm1, epta_wm2, epta_wm3, epta_interpret}),
we model it as a spatially uncorrelated red noise process.
Individual Fourier components of this broadband signal may contaminate frequencies at which the presence of ULDM is evaluated.
Thus, this common red  noise signal is included in the null hypothesis, $\varnothing$, along with the pulsar-intrinsic noise.
The signal hypothesis is based on adding ULDM to the null hypothesis.

\begin{figure*}[!htb]
    \centering
    \begin{subfigure}[b]{0.49\textwidth}
        \includegraphics[width=\textwidth]{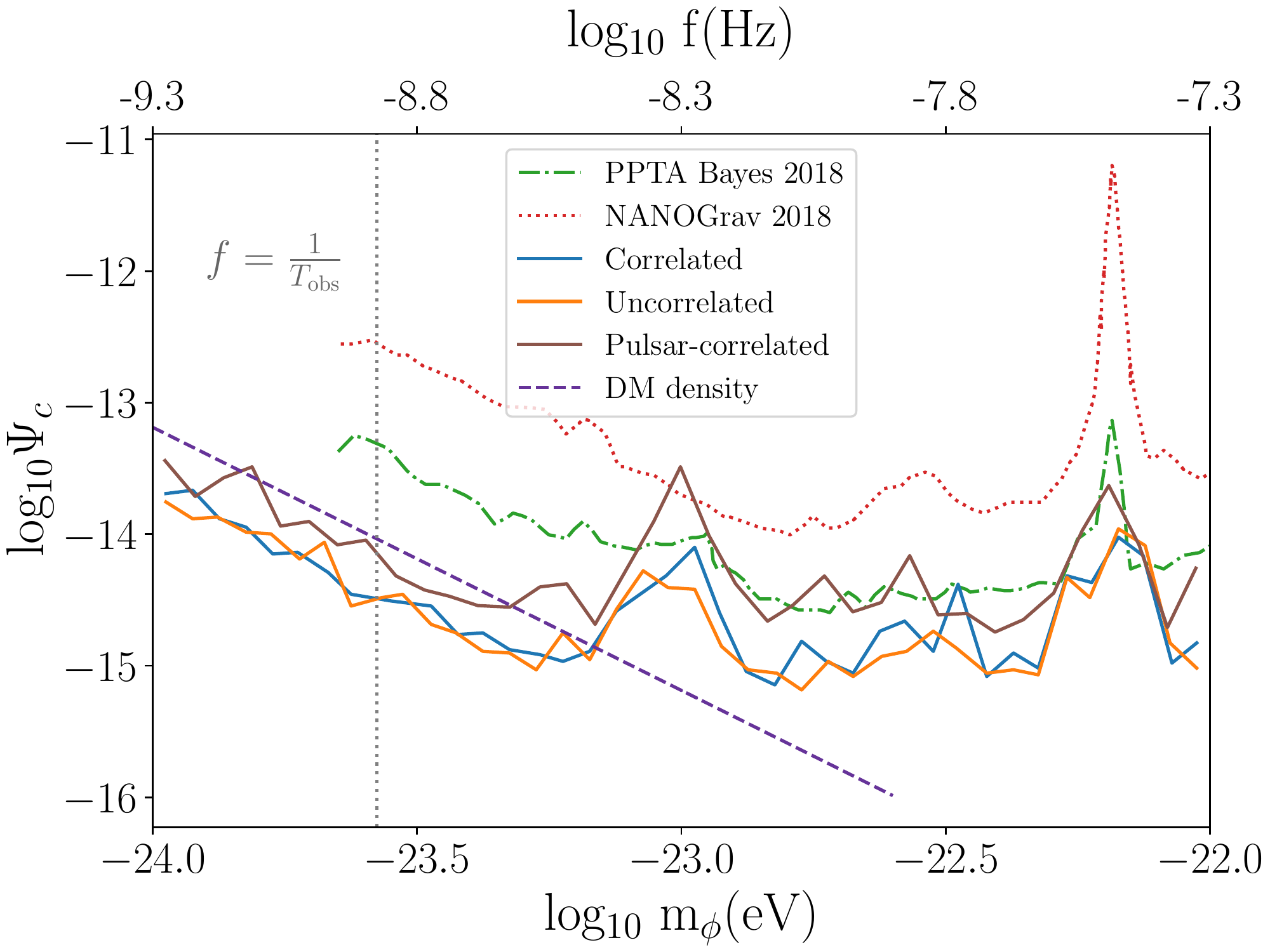}
        \label{fig:pe_xg:m}
    \end{subfigure}
    \begin{subfigure}[b]{0.49\textwidth}
        \includegraphics[width=\textwidth]{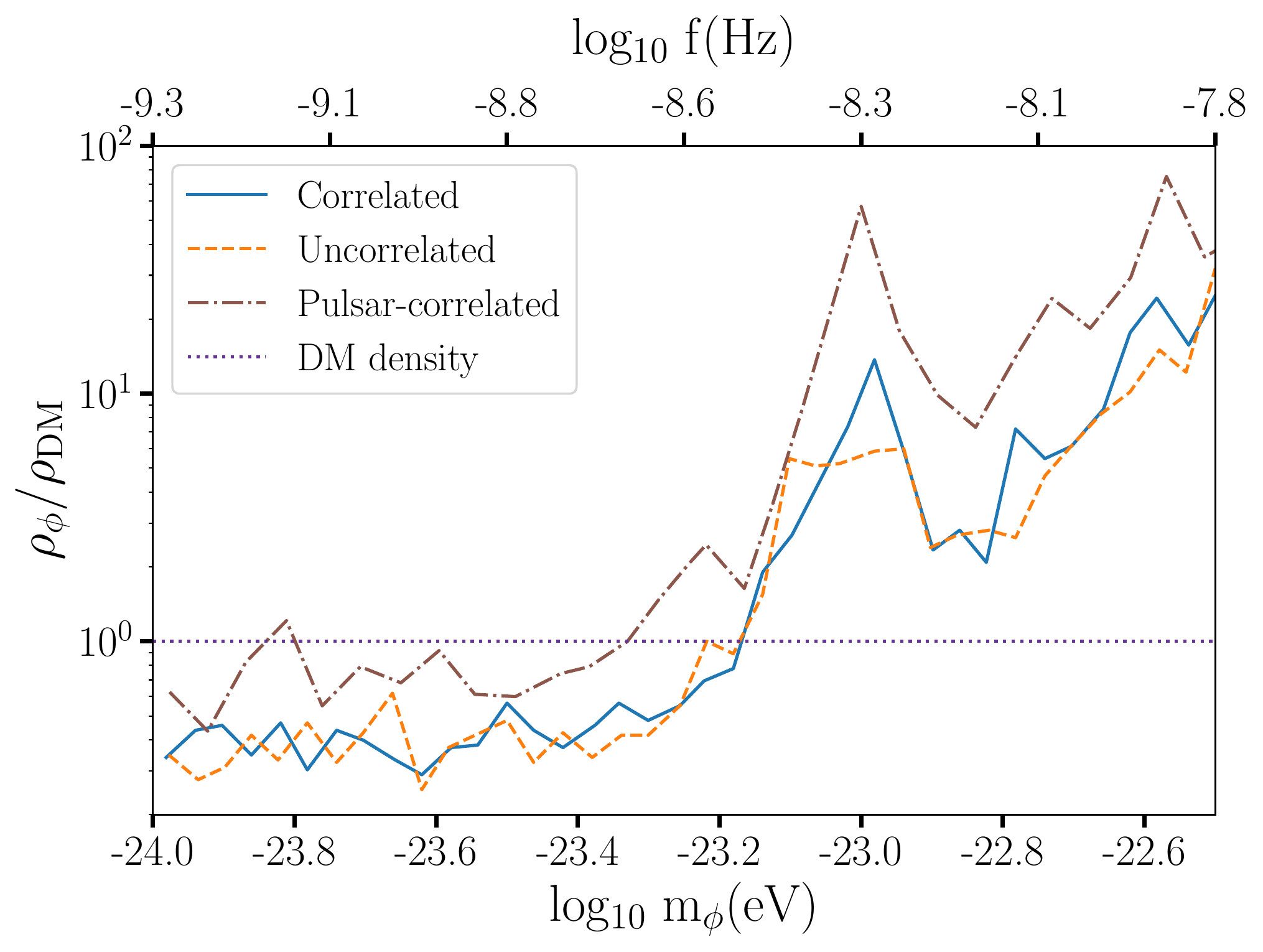}
        \label{fig:pe_xg:nom}
    \end{subfigure}
    \vspace{-0.5\baselineskip} 
    \caption{Upper limits on  ULDM, and namely on the dimensionless amplitude ($\Psi_\text{c}$, \textit{left panel}) and the ULDM fraction of the local DM density $\rho_\text{DM} = 0.4 \text{GeV}/\text{cm}^3$ ($\rho_\phi / \rho_{\text{DM}}$, \textit{right panel}), at 95\% credibility. The bottom horizontal axes show the ULDM particle mass, whereas the top horizontal axes show the equivalent oscillation frequency of the scalar field.
    The upper limits from previous searches \cite{Porayko_2014,Porayko_2018} are shown for comparison.
    As a reference, we highlight the frequency  $T_\text{obs}^{-1}$. 
    In the right panel, we zoom in on the excluded ULDM masses. The horizontal dotted line represents the value of $\rho_\phi$ that would saturate the local DM density.
    Notice that based on our results ULDM particles with mass $-24.0 < \text{log}_{10}\text{~}(m_\phi/\text{eV}) < -23.7$ can only make up at most $30- 40$ \% of the total DM energy density, while particles with mass $-23.7< \text{log}_{10}\text{~}(m_\phi/\text{eV}) < -23.3$ can contribute at most up to  $ \sim 70$ \%.
    } 
    \label{fig:25yrULDM}
\end{figure*}

\textit{Results.}---We carry out the search for ULDM in the correlated, pulsar-correlated and uncorrelated limit for $\hat{\phi}^2_\text{E}$ and $\hat{\phi}^2_\text{P}$ with the parameters in Table~\ref{tab:priors}. The factors $\hat{\phi}^2_\text{E}$ and $\hat{\phi}^2_\text{P}$ are drawn from an exponential prior, to correctly model the stochastic nature of the ULDM field~\cite{Foster_2018, Centers_2021}.  
We find no evidence for a signal in the mass range  $m_\phi \sim [10^{-24}~\text{eV}, 10^{-22}~\text{eV}]$.
The largest $\ln\mathcal{B}$ we find across frequency-mass bins is $<1$, i.e. the null hypothesis is favored.
Therefore, we calculate the 95\% upper limits on the signal amplitude $\Psi_\text{c}$ and, through Eq.~\eqref{eq:psi_c}, on the scalar field density $\rho_\phi$.
The results are shown in Fig.~\ref{fig:25yrULDM}.

A highlight of our study is that not only does EPTA DR2 yield more stringent constraints than previous results~\cite{Porayko_2014, Porayko_2018}, but it also rules out that particles with masses $m_\phi \sim [10^{-24}~\text{eV}, 10^{-23.3}~\text{eV}]$ can be  100\% of the observed local DM density. In particular,  the scalar field density is $\rho_\phi \lesssim 0.15~\text{GeV}/\text{cm}^3$ in the mass range $m_\phi \sim [10^{-24}~\text{eV}, 10^{-23.7}~\text{eV}]$, while it is constrained to $\rho_\phi \lesssim 0.30~\text{GeV}/\text{cm}^3$ between $m_\phi \sim [10^{-23.7}~\text{eV}, 10^{-23.4}~\text{eV}]$ .
Furthermore, the correlated limit in Fig.~\ref{fig:25yrULDM} confirms Lyman-$\alpha$ bounds, which exclude ULDM in this mass range unless it constitutes less than $30\%$ of DM \cite{Kobayashi_2017}. It is worth noticing that the low-frequency end of Fig.~\ref{fig:25yrULDM} extends below the n{\"a}ive expectation $f = 1.3~\text{nHz}$ corresponding to the inverse of the observation time $T_\text{obs} = 24.7~\text{yr}$. In fact, while an ULDM candidate in this mass region does not complete an oscillation cycle during the observation timescale, the signal can still be approximated by a polynomial expansion in $(m_\phi t)$ \citep{Kaplan_2022}.
The sensitivity in this region is limited by the simultaneous fitting to pulsar spin frequency derivatives~\cite{HazbounRomano2019,BlandfordNarayan1984}.
PTAs are only sensitive to the $(m_\phi t)^3$ term, as the first terms in the expansion are degenerate with the timing model \citep{Ramani_2020}. However, since the expected amplitude $\Psi_\text{c}$ of an ULDM candidate increases as its mass decreases, we can still obtain competitive constraints at low frequency. While, in principle, our analysis could be pushed to even lower masses~\cite{Unal_2022}, we choose to focus on the region $m_\phi \gtrsim 10^{-24}~\text{eV}$ to comply with the aforementioned CMB bounds.
We find that the significant improvement in sensitivity to ULDM at low frequencies arises thanks to the larger data span of EPTA DR2, in accordance with the theoretical sensitivity scaling proposed in Eq.~(13) of Ref.~\cite{Unal_2022}. 
In particular, because of the longer data span, we expect EPTA DR2 limits  to  be  better  than  NANOGrav \cite{Afzal_2023} ones  by  a  factor of roughly $\sim 3.6$,  which is in agreement with what observed.
At high frequencies, we find that the advantage of the long timing baseline compared to NANOGrav diminishes, also in accordance with the scaling, as pulsar white noise levels become more important.
We also performed an identical analysis of the 10-year subset of EPTA DR2~\cite{epta_wm1,epta_wm3}, as well as of the MeerTime data~\cite{Miles_2022}, which yield less stringent upper limits in agreement with the scaling. For comparison, the bounds in both the correlated and  uncorrelated limit for the 10-year subset of the EPTA, shown in Fig.~22 in Ref.~\cite{epta_interpret}, appear at the level of the pulsar-correlated limit in Fig.~\ref{fig:25yrULDM} of this manuscript.
\interfootnotelinepenalty=10000

In the following, we clarify some specific aspects of our results.
First, we notice that a similar analysis has been done by the North American Nanohertz Observatory for Gravitational Waves (NANOGrav) collaboration~\cite{Afzal_2023}.
There, the upper limits provided in the correlated and uncorrelated scenarios differ at low frequency.
This can be understood by noticing that the correlated limit of NANOGrav corresponds to our pulsar-correlated limit.
However, in the low mass limit of Fig.~\ref{fig:25yrULDM}, the pulsars, the Earth and the stellar and gaseous tracers used for rotation curves estimates lie well within the area spanned by the coherence length; thus, one can only measure the combination $\Psi_c^0 =  \Psi_c \hat\phi^2$, which represents the realization of DM in our Galaxy.
Therefore, we remove the $\hat\phi^2_\text{E} = \hat\phi^2_\text{P}\equiv \hat\phi^2$ parameter in the correlated limit, as it can be accounted for by a redefinition of $\Psi_c$. 
Fitting for $\Psi_c$ and $\hat\phi^2$ separately, instead, introduces an additional uncertainty, which leads our pulsar-correlated analysis to produce a similar mismatch as the one found in Ref.~\cite{Afzal_2023}, as shown in Fig.~\ref{fig:25yrULDM}. 
Second, Fig.~\ref{fig:25yrULDM} hints at a steep increase in the upper limits at $m_\phi \gtrsim 10^{-23.2}~\text{eV}$. In fact, we report the presence of excess signal power on top of the common red noise process, corresponding to a mass of $m_\phi \simeq 10^{-23}~\text{eV}$ and an amplitude of $\Psi_\text{c} \simeq 6 \times 10^{-14}$ or, equivalently, a density of $\rho_\phi = 90~\text{GeV}/\text{cm}^3$. At face value, this excess is not compatible with an ULDM candidate, as  the corresponding density is outside the local DM measurement uncertainties \cite{Bovy_2012, Read_2014, Sivertsson_2018, de_Salas_2020}. Moreover, such a mass would be in tension with astrophysical bounds, as extensively discussed in the introduction \cite{Hlozek_2015, Irsic_2017, Armengaud_2017, Kobayashi_2017, Nori_2018, Rogers_2021, Hayashi_2021, Dalal_2022}.
Anyway, the Bayesian odds ratio suggests that it is still consistent with noise ($\ln\mathcal{B} \sim 0.1$). We find a similar excess in the analysis of 10-yr subset of the EPTA DR2~\cite{epta_interpret}. Moreover, the boson mass corresponding to the excess also matches the frequency of the continuous gravitational wave (CGW) candidate studied in \cite{epta_wm4}. This motivates further investigations as part of the International Pulsar Timing Array~\cite{VerbiestLentati2016}.

\textit{Conclusions.}---
ULDM is a theoretically motivated paradigm that may alleviate the \textit{small-scale crisis} of structure formation. Here, we focused on the most robust scenario, in which ULDM features only gravitational interactions.
These interactions produce a periodic oscillation in the TOAs of  the radio beams emitted by pulsars, which can then be collected in PTA telescopes.
PTAs stand out as excellent laboratories to test the effects of ULDM in the mass range $m_\phi \sim [10^{-24}~\text{eV}, 10^{-22}~\text{eV}]$. In this work, we showed that PTAs constrain the presence of ULDM \textit{below} a few tenths of the observed DM abundance in the mass range $m_\phi \sim [10^{-24}~\text{eV}, 10^{-23.3}~\text{eV}]$. 
Therefore, in this range, ULDM cannot constitute 100\% of the observed DM.

We wish to thank Diego Blas, Alessio Zicoschi and Paolo Salucci for useful discussions and insights. Moreover, we thank the anonymous referees for improving the quality of our manuscript. 
C.S. and E.B. acknowledge support from the European Union’s H2020 ERC Consolidator Grant “GRavity from Astrophysical to Microscopic Scales” (Grant No. GRAMS-815673) and the EU Horizon 2020 Research and Innovation Programme under the Marie Sklodowska-Curie Grant Agreement No. 101007855.
B.G. is supported by the Italian Ministry of Education, University and Research within the PRIN 2017 Research Program Framework, n. 2017SYRTCN.
The European Pulsar Timing Array (EPTA) is a collaboration between
European and partner institutes, namely ASTRON (NL), INAF/Osservatorio
di Cagliari (IT), Max-Planck-Institut f\"{u}r Radioastronomie (GER),
Nan\c{c}ay/Paris Observatory (FRA), the University of Manchester (UK),
the University of Birmingham (UK), the University of East Anglia (UK),
the University of Bielefeld (GER), the University of Paris (FRA), the
University of Milan-Bicocca (IT), the Foundation for Research and 
Technology, Hellas (GR), and Peking University (CHN), with the
aim to provide high-precision pulsar timing to work towards the direct
detection of low-frequency gravitational waves. An Advanced Grant of
the European Research Council allowed to implement the Large European Array
for Pulsars (LEAP) under Grant Agreement Number 227947 (PI M. Kramer). 
The EPTA is part of the
International Pulsar Timing Array (IPTA); we thank our
IPTA colleagues for their support and help with this paper and the external Detection Committee members for their work on the Detection Checklist.
The work presented here is a culmination of many years of data
analysis as well as software and instrument development. In particular,
we thank Drs. N.~D'Amico, P.~C.~C.~Freire, R.~van Haasteren, 
C.~Jordan, K.~Lazaridis, P.~Lazarus, L.~Lentati, O.~L\"{o}hmer and 
R.~Smits for their past contributions. We also
thank Dr. N. Wex for supporting the calculations of the
galactic acceleration as well as the related discussions.
The EPTA is also grateful to staff at its observatories and telescopes who have made the continued observations possible.
Part of this work is based on observations with the 100-m telescope of
the Max-Planck-Institut f\"{u}r Radioastronomie (MPIfR) at Effelsberg
in Germany. Pulsar research at the Jodrell Bank Centre for
Astrophysics and the observations using the Lovell Telescope are
supported by a Consolidated Grant (ST/T000414/1) from the UK's Science
and Technology Facilities Council (STFC). ICN is also supported by the
STFC doctoral training grant ST/T506291/1. The Nan{\c c}ay radio
Observatory is operated by the Paris Observatory, associated with the
French Centre National de la Recherche Scientifique (CNRS), and
partially supported by the Region Centre in France. We acknowledge
financial support from ``Programme National de Cosmologie and
Galaxies'' (PNCG), and ``Programme National Hautes Energies'' (PNHE)
funded by CNRS/INSU-IN2P3-INP, CEA and CNES, France. We acknowledge
financial support from Agence Nationale de la Recherche
(ANR-18-CE31-0015), France. The Westerbork Synthesis Radio Telescope
is operated by the Netherlands Institute for Radio Astronomy (ASTRON)
with support from the Netherlands Foundation for Scientific Research
(NWO). The Sardinia Radio Telescope (SRT) is funded by the Department
of University and Research (MIUR), the Italian Space Agency (ASI), and
the Autonomous Region of Sardinia (RAS) and is operated as a National
Facility by the National Institute for Astrophysics (INAF).
The work is supported by the National SKA programme of China
(2020SKA0120100), Max-Planck Partner Group, NSFC 11690024, CAS
Cultivation Project for FAST Scientific. This work is also supported
as part of the ``LEGACY'' MPG-CAS collaboration on low-frequency
gravitational wave astronomy. 
J.A. acknowledges support from the
European Commission (Grant Agreement number: 101094354). J.A. and S. Chanlaridis 
were partially supported by the Stavros
Niarchos Foundation (SNF) and the Hellenic Foundation for Research and
Innovation (H.F.R.I.) under the 2nd Call of the ``Science and Society --
Action Always strive for excellence -- Theodoros Papazoglou''
(Project Number: 01431). A.C. acknowledges support from the Paris
\^{I}le-de-France Region. A.C., A.F., A. Sesana, A. Samajdar, E.B., D.I., G.M.S., M. Bonetti acknowledge
financial support provided under the European Union's H2020 ERC
Consolidator Grant ``Binary Massive Black Hole Astrophysics'' (B
Massive, Grant Agreement: 818691). G.D., K. Liu, R.K. and M.K. acknowledge support
from European Research Council (ERC) Synergy Grant ``BlackHoleCam'', 
Grant Agreement Number 610058. This work is supported by the ERC 
Advanced Grant ``LEAP'', Grant Agreement Number 227947 (PI M. Kramer). 
A.V. and P.R.B. are supported by the UK's Science
and Technology Facilities Council (STFC; grant ST/W000946/1). A.V. also acknowledges
the support of the Royal Society and Wolfson Foundation. N.K.P. is funded by the Deutsche
Forschungsgemeinschaft (DFG, German Research Foundation) --
Projektnummer PO 2758/1--1, through the Walter--Benjamin
programme. A. Samajdar thanks the Alexander von Humboldt foundation in
Germany for a Humboldt fellowship for postdoctoral researchers. A. Possenti, D.P.
and M. Burgay acknowledge support from the research grant “iPeska”
(P.I. Andrea Possenti) funded under the INAF national call
Prin-SKA/CTA approved with the Presidential Decree 70/2016
(Italy). R.N.C. acknowledges financial support from the Special Account
for Research Funds of the Hellenic Open University (ELKE-HOU) under
the research programme ``GRAVPUL'' (K.E.-80383/grant agreement 319/10-10-2022: PI N. A. B.
Gizani).
E. vd. W., C.G.B. and G.H.J. acknowledge support from the Dutch National Science
Agenda, NWA Startimpuls – 400.17.608.
L.S. acknowledges the use of the HPC system Cobra at the Max Planck Computing and Data Facility.
\appendix



\providecommand{\noopsort}[1]{}\providecommand{\singleletter}[1]{#1}%

\end{document}